\documentclass[aps,prc,twocolumn,floatfix,nofootinbib,showpacs,superscriptaddress]{revtex4-1}
\usepackage{amsmath,bm}
\usepackage{enumerate}
\usepackage{graphicx}
\usepackage{subfigure}
\usepackage{float}
\usepackage{longtable}
\usepackage{color}

\begin{document}

%\title{Explaining masses of ground-state light-quark mesons}
\title{Tracing masses of ground-state light-quark mesons}

\author{Lei Chang}%(³£À×)}
\affiliation{Department of Physics, Center for High Energy Physics and State Key Laboratory of Nuclear Physics and Technology, Peking University, Beijing 100871,
China}
\affiliation{Physics Division, Argonne National Laboratory, Argonne,
Illinois 60439, USA}

\author{Craig D. Roberts}
%\email[Corresponding author: ]{cdroberts@anl.gov}
\affiliation{Department of Physics, Center for High Energy Physics and State Key Laboratory of Nuclear Physics and Technology, Peking University, Beijing 100871,
China}
\affiliation{Physics Division, Argonne National Laboratory, Argonne,
Illinois 60439, USA}

%\date{\today}

\begin{abstract}
We describe a symmetry-preserving calculation of the light-quark meson spectrum, which combines a description of pion properties with reasonable estimates of the masses of heavier mesons, including axial-vector states.  The kernels used in formulating the problem are essentially nonperturbative, incorporating effects of dynamical chiral symmetry breaking (DCSB) that were not previously possible to express.  Our analysis clarifies a causal connection between DCSB and the splitting between vector and axial-vector mesons, and exposes a key role played by the anomalous chromomagnetic moment of dressed-quarks in forming the spectrum.
%
%\vspace*{13ex}
%
\end{abstract}

\pacs{
11.10.St, 	%Bound and unstable states; Bethe–Salpeter equations
11.30.Rd, 	%Chiral symmetries
12.38.Lg,   %Other nonperturbative calculations
24.85.+p 	%Quarks, gluons, and QCD in nuclear reactions
}

\maketitle

\section{Prologue}
Spectroscopy is basic to exposing the character of Nature's fundamental forces, e.g.: the hydrogen atom spectrum is intimately connected with the development of quantum mechanics and quantum electrodynamics; and cataloguing the surfeit of hadrons -- the pion- and proton-like particles, initially supposed to be elementary -- led to the notion of quarks, with their fractional electric charge, and ultimately to quantum chromodynamics (QCD), the strongly-interacting piece of the Standard Model.
%Spectroscopy has long been a basic tool in our progress toward exposing the character of Nature's fundamental forces.  For example: the hydrogen atom spectrum is intimately connected with the development of quantum mechanics and quantum electrodynamics; and cataloguing the plethora of hadrons -- the pion- and proton-like particles, initially supposed to be elementary -- led to the notion of quarks, with their fractional electric charge, and ultimately to quantum chromodynamics (QCD), the strongly-interacting piece of the Standard Model.

QCD is peculiar owing to the empirical fact of \emph{confinement}; namely, whilst it is formulated in terms of quarks and gluons, which carry the color quantum number, these degrees of freedom have never been directly observed.  Determination of the spectrum of color-singlet hadrons is therefore one of the few ways by which to elucidate the long-range behavior of the interaction between quarks and gluons, and thereby grasp the essence of Nature's sole confining force.  It might thus be surprising that more than thirty years after the discovery of the first light-quark meson with mass greater than 1\,GeV, there is no reliable, symmetry-preserving computation of the ground-state spectrum of light-quark mesons.  This is particularly disturbing given the importance of spectroscopy at next-generation hadron physics facilities \cite{Carman:2005ps}.

Confinement is much misapprehended.  We therefore remark that the static potential measured in lattice-QCD simulations is not related in any known way to the puzzle of light-quark confinement.  Light-quark creation and annihilation effects are fundamentally nonperturbative.  It is thus impossible in principle to compute a potential between light quarks.  Alternatively, arguments relating confinement to the analytic properties of QCD's Schwinger functions have been presented \cite{Krein:1990sf,Roberts:2007ji}, from which perspective the question of light-quark confinement may be translated into the challenge of charting the infrared behavior of QCD's \emph{universal} $\beta$-function.  Solving this problem is a basic goal of modern physics, which can be addressed in any framework that enables the nonperturbative evaluation of renormalization constants.

%Regarding confinement, little is known and much is misapprehended.  It is therefore important to state clearly that the static potential measured in simulations of lattice-QCD is not related in any known way to the question of light-quark confinement.  It is a basic feature of QCD that light-quark creation and annihilation effects are fundamentally nonperturbative.  Hence it is impossible in principle to compute a potential between two light quarks.  On the other hand, arguments relating confinement to the analytic properties of QCD's Schwinger functions have been presented \cite{Roberts:2007ji}, from which perspective the question of light-quark confinement may be translated into the challenge of charting the infrared behavior of QCD's \emph{universal} $\beta$-function.  Solving this well-posed problem is an elemental goal of modern physics.%, which can be addressed in any framework that enables the nonperturbative evaluation of renormalization constants.

The light-hadron spectrum has challenged lattice-QCD since its formulation almost forty years ago \cite{Wilson:2004de}.  The approach involves a range of systematic errors, amongst which explicit chiral symmetry breaking is particularly damaging to studies of light-quark systems.  It has only recently become possible to tame these errors and reproduce the mass and width of the $\rho$-meson \cite{Durr:2008zz}.  However, in connection with more massive light-quark mesons, available results are much affected by lattice artifacts \cite{Dudek:2010wm,Engel:2010my}.

%The light-hadron spectrum has challenged lattice-QCD since its formulation more than thirty-five years ago \cite{Wilson:2004de}.  The approach involves systematic errors associated with: discretization; finite volume; too-large dynamical quark masses; and violation of symmetries, amongst which the explicit breaking of chiral symmetry is particularly damaging for studies of light-quark systems.  Only recently has it become possible to control these errors so that the known mass and width of the $\rho$-meson could accurately be reproduced \cite{Durr:2008zz}.  However, in connection with more massive light-quark mesons, the available results are much affected by lattice artifacts \cite{Dudek:2010wm,Engel:2010my}.

We prefer a continuum nonperturbative approach and, exploiting a novel form of the Bethe-Salpeter equation (BSE) \cite{Chang:2009zb}, formulate a tractable, symmetry preserving expression of the meson bound-state problem.  This treatment connects the $\beta$-function to observables, so that comparison between computations and the spectrum may be used to constrain the $\beta$-function's long-range behavior, as illustrated in Ref.\,\cite{Qin:2011dd}.

\section{Bound-State Equations}
Our approach begins with the gap equation %($f$ labels the quark flavor):%\footnote{In our Euclidean metric:  $\{\gamma_\mu,\gamma_\nu\} = 2\delta_{\mu\nu}$; $\gamma_\mu^\dagger = \gamma_\mu$; $\gamma_5= \gamma_4\gamma_1\gamma_2\gamma_3$; $a \cdot b = \sum_{i=1}^4 a_i b_i$; and $P_\mu$ timelike $\Rightarrow$ $P^2<0$.}
\begin{eqnarray}
\nonumber S_f(p)^{-1} & = & Z_2 \,(i\gamma\cdot p + m_f^{\rm bm})+ Z_1 \int^\Lambda_{dq}\!\! g^2 D_{\mu\nu}(p-q) \\
&& \times \frac{\lambda^a}{2}\gamma_\mu S_f(q) \frac{\lambda^a}{2}\Gamma^f_\nu(q,p) ,
\label{gendse}
\end{eqnarray}
where: $f$ labels quark flavor; $D_{\mu\nu}$ is the gluon propagator; $\Gamma^f_\nu$, the quark-gluon vertex; $\int^\Lambda_{dq}$, a Poincar\'e invariant regularization of the integral, with $\Lambda$ the regularization scale; $m^{\rm bm}(\Lambda)$, the current-quark bare mass; and $Z_{1,2}(\zeta,\Lambda)$, respectively, the vertex and quark wave function renormalization constants, with $\zeta$ the renormalization point -- dependence upon which we do not usually make explicit.  The solution is the quark propagator,
\begin{eqnarray}
%\nonumber
 S_f(p)^{-1} & = & i \gamma\cdot p \, A_f(p^2) + B_f(p^2) \,.%\\
%
%& =& \frac{1}{Z(p^2,\zeta^2)}\left[ i\gamma\cdot p + M(p^2,\zeta^2)\right] .
\label{sinvp}
\end{eqnarray}
It is obtained from Eq.\,(\ref{gendse}) augmented by a renormalization condition.

The mass of all mesons with the same quantum numbers may be obtained from a single inhomogeneous BSE.  The dressed-quark propagator is crucial in constructing its kernel.  The gap equation involves the dressed-quark gluon vertex, $\Gamma_\mu$.  Following Ref.\,\cite{Chang:2009zb}, one can now construct a symmetry-preserving kernel for the Bethe-Salpeter kernel given any form for $\Gamma_\mu$.
%Owing to the importance of symmetries in forming the spectrum of a quantum field theory, this is a pivotal advance.
Hence, in solving bound-state problems one is no longer reliant upon step-by-step improvements in the computation of $\Gamma_\mu$ \cite{Bender:1996bb,Watson:2004kd,Fischer:2009jm}; instead, one may use all available, reliable information to construct the best possible \emph{Ansatz}.  This enables one to incorporate crucial nonperturbative effects, which any finite sum of contributions is incapable of capturing.  In this way the nonperturbative phenomenon of dynamical chiral symmetry breaking (DCSB) was shown to generate material, momentum-dependent anomalous chromo- and electro-magnetic moments for dressed light-quarks \cite{Chang:2010hb}.

DCSB is a keystone of the Standard Model.  This remarkable mass-generating mechanism can be explained via $S(p)$.  In the chiral limit, the mass function, $M(p^2) = B(p^2)/A(p^2)$, is identically zero at any finite order in perturbation theory.  However, DCSB generates mass from \emph{nothing}.  Thus, in chiral-QCD, despite the absence of an explicit mass source, $M(p^2)$ is nonzero and strongly momentum-dependent, with $M(0) \approx 0.5\,$GeV \cite{Bhagwat:2003vw,Bowman:2005vx}.  DCSB is responsible for constituent-quark masses and intimately connected with confinement but it is not known whether the connection is accidental or causal.  Nevertheless, DCSB is the most important mass-generating mechanism for visible matter in the Universe.  It generates roughly 98\% of a proton's mass and amplifies the Higgs mechanism for explicit symmetry breaking \cite{Flambaum:2005kc}.

Herein we calculate masses of ground-state spin-zero and -one light-quark mesons, in order to illuminate the impact of DCSB on the spectrum.  Given its central role, we explicate the problem in the pseudoscalar, $J^{PC}=0^{-+}$, and axial-vector, $1^{++}$, channels and simply indicate the changes required to study $0^{++}$, $1^{--}$, $1^{+-}$.  %Results for tensor, hybrid and exotic states will be reported elsewhere.

Pseudoscalar and axial-vector mesons appear as poles in the inhomogeneous Bethe-Salpeter amplitude associated with the axial-vector vertex, $\Gamma_{5\mu}^{fg}$.  An exact form of the associated BSE is ($q_\pm=q\pm P/2$, etc.)
% Lambda^0 -> Z_1^{-1} Lambda
\begin{eqnarray}
\nonumber
\Gamma_{5\mu}^{fg}(k;P) & = & Z_2 \gamma_5\gamma_\mu - Z_1\!\int_{dq} g^2D_{\alpha\beta}(k-q)\frac{\lambda^a}{2}\,\gamma_\alpha S_f(q_+) \\
\nonumber
&&  \times \Gamma_{5\mu}^{fg}(q;P) S_g(q_-) \frac{\lambda^a}{2}\,\Gamma_\beta^g(q_-,k_-) \\
\nonumber
&& + Z_1 \! \int_{dq} g^2D_{\alpha\beta}(k-q)\, \frac{\lambda^a}{2}\,\gamma_\alpha S_f(q_+)\\
&& \times  \frac{\lambda^a}{2} \Lambda_{5\mu\beta}^{fg}(k,q;P), \label{genbse}
\end{eqnarray}
where $\Lambda_{5\mu\beta}^{fg}$ is a four-point function, completely defined \cite{Chang:2009zb} via the quark self-energy and hence the quark-gluon vertex, $\Gamma_\mu$.  Crucially, $\Lambda_{5\mu\beta}^{fg}$ satisfies a Ward-Takahashi identity, whose solution provides a symmetry-preserving \emph{Ansatz} consistent with $\Gamma_\mu$.  Bound-state masses in the $0^{++}$, $1^{--}$, $1^{+-}$ channels are, respectively, obtained by replacing $Z_2 \gamma_5\gamma_\mu$ by the structures $I_{\rm D}$, $\gamma_\mu$, $\gamma_5 k_\mu$, each multiplied by an appropriate renormalization constant.

%The mass of the ground-state in a given channel can reliably be obtained using the method detailed in Ref.\,\cite{Bhagwat:2007rj}.  Namely, one: solves the inhomogeneous BSE on a large set of points at spacelike total momenta, $P^2$; focuses on the dominant amplitude in the channel, which for the pseudoscalar is $i \gamma_5 E_5(k;P)$, with $E_5$ a scalar function; and locates the zero at timelike momentum in $[1/E_5(k=0;P)]$, or its analog in other channels, via analysis with Pad\'e approximants.  This is an equivalent of the method used to compute masses in lattice-QCD.

A prediction for the spectrum therefore follows once the gap equation's kernel is specified and the Ward-Identity solved for $\Lambda_{5\mu\beta}^{fg}$.  The kernel may be rendered tractable by writing \cite{Maris:1997tm,Bloch:2002eq}
\begin{equation}
Z_1 g^2 D_{\rho \sigma}(t) \Gamma_\sigma(q,q+t)
= \tilde{\cal G}(t^2) \, D_{\rho\sigma}^{\rm free}(t) Z_2 \tilde\Gamma_\sigma(q,q+t)\,, \label{KernelAnsatz}
\end{equation}
wherein $D_{\rho \sigma}^{\rm free}$ is the Landau-gauge free-gauge-boson pro\-pa\-ga\-tor, $\tilde{\cal G}$ is an interaction model and $\tilde\Gamma_\sigma$ is an \emph{Ansatz} for the quark-gluon vertex.  For the interaction, we employ
%\begin{equation}
%\label{IRGs}
%{\cal G}(t^2) = \frac{4\pi^2}{\omega^6}\, D\, t^4\, {\rm e}^{-t^2/\omega^2},
%\end{equation}
\begin{equation}
\label{CalGQC}
\frac{\tilde{\cal G}(s)}{s} = \frac{8 \pi^2}{\omega^4} D \, {\rm e}^{-s/\omega^2}
+ \frac{8 \pi^2 \gamma_m\, {\cal F}(s)}{\ln [ \tau + (1+s/\Lambda_{\rm QCD}^2)^2]} ,
\end{equation}
where: $\gamma_m = 12/25$, $\Lambda_{\rm QCD}=0.234\,$GeV; $\tau={\rm e}^2-1$; and ${\cal F}(s) = \{1 - \exp(-s/[4 m_t^2])\}/s$, $m_t=0.5\,$GeV.  This form preserves the one-loop renormalization-group behavior of QCD in the gap- and Bethe-Salpeter-equations, and is consonant with modern DSE- and lattice-QCD results.  Detailed explanations of its development and capability as a tool in hadron physics are presented in Refs.\,\cite{Qin:2011dd,Qin:2011xq}.  We employ the renormalization procedures of Ref.\,\cite{Maris:1997tm} and the same renormalization point, $\zeta=19\,$GeV.  N.B.\ This study is the first to employ the fully-renormalized forms of the vertex-dressed gap equation and the symmetry-preserving Bethe-Salpeter equations that are derived from it.

We adapt the vertex explained in Refs.\,\cite{Chang:2010hb,Ball:1980ay,Bashir:2011dp}, viz.:
\begin{eqnarray}
\label{ourvtx}
\tilde\Gamma_\mu(p_1,p_2)  & = & \Gamma_\mu^{\rm BC}(p_1,p_2) +
\Gamma_\mu^{\rm acm}(p_1,p_2)\,;\\
\nonumber
i\Gamma_\mu^{\rm BC}(p_1,p_2)  & = &
i\Sigma_A(p_1^2,p_2^2)\,\gamma_\mu + 2 \ell_\mu \left[ i\gamma\cdot \ell \,\Delta_A(p_1^2,p_2^2)  \right. \\
&&  \left. + \Delta_B(p_1^2,p_2^2)\right] ,
\label{bcvtx}
\end{eqnarray}
where $\Sigma_{\phi}(p_1^2,p_2^2) = [\phi(p_1^2)+\phi(p_2^2)]/2$, $\Delta_{\phi}(p_1^2,p_2^2) = [\phi(p_1^2)-\phi(p_2^2)]/[p_1^2-p_2^2]$, $2 \ell = p_1+p_2$; and the anomalous chromomagnetic moment piece is
\begin{equation}
\Gamma_\mu^{\rm acm}(p_1,p_2) = \Gamma_\mu^{\rm acm_4}(p_1,p_2) + \Gamma_\mu^{\rm acm_5}(p_1,p_2)\,,
\end{equation}
with ($k=p_1-p_2$, $T_{\mu\nu} = \delta_{\mu\nu} - k_\mu k_\nu/k^2$, $a_\mu^{\rm T} := T_{\mu\nu}a_\nu$)
\begin{eqnarray}
\Gamma_\mu^{\rm acm_4} &=& [ \ell_\mu^{\rm T} \gamma\cdot  k + i \gamma_\mu^{\rm T} \sigma_{\nu\rho}\ell_\nu k_\rho] \tau_4(p_1,p_2)\,,\\
%
%\nonumber
%
\Gamma_\mu^{\rm acm_5} & =& \sigma_{\mu\nu}k_\nu\tau_5(p_1,p_2)\,,\\
\tau_4 &=& \frac{2 \tau_5(p_1,p_2)}{\mathcal{M}(p_1^2,p_2^2)}\,,
%\bigg[  \frac{4}{{\cal M}(p_1^2,p_2^2)}\Delta_B(p_1^2,p_2^2) - \Delta_A(p_1^2,p_2^2) \bigg],
\label{tau4}
\end{eqnarray}
$\tau_5 =  \eta\, \Delta_B(p_1^2,p_2^2)$ and ${\cal M}(x,y)=[x+M(x)^2+y+M(y)^2]/(2[M(x)+M(y)])$.
%and $\Gamma_\mu^{\rm acm_5}=\sigma_{\mu\nu}k_\nu\tau_5(p_1,p_2)$,
%

Our \emph{Ansatz} combines information and results from perturbative QCD, Dyson-Schwinger equation (DSE) and lattice-QCD studies.
The structure and importance of the BC component has long been recognized \cite{Roberts:1994dr}: its presence and momentum dependence are confirmed in comparisons between DSE- and lattice-QCD studies \cite{Skullerud:2003qu,Bhagwat:2004kj,Skullerud:2004gp}.
The ACM piece is novel, although its presence should long ago have been appreciated, given the importance of the DCSB-induced $\Delta_B$-term in the BC component and the ACM term's identical origin.

Our introduction of the ACM term comes at a critical juncture, since the developments in Ref.\,\cite{Chang:2009zb} only now enable an exploration of its consequences for hadron observables.
Indeed, notwithstanding the large current-quark mass employed (115\,MeV), a comparison between lattice-QCD estimates of $\tau_5(p_1,p_2)$ \cite{Skullerud:2003qu,Skullerud:2004gp} and our expression confirms the character of our \emph{Ansatz}.  For example, there is a two orders-of-magnitude enhancement over the perturbative result, emphasizing the connection with DCSB, and semi-quantitative agreement between the momentum-dependence of the lattice result and our form.
It is notable that in perturbation theory, whilst the on-shell anomalous chromomagnetic moment is always negative for a dressed-quark, the sign of the $\tau_5(p_1,p_2)$ contribution is positive in Landau gauge but negative in Feynman gauge \cite{Davydychev:2000rt}.
To complete the picture, a further consideration of perturbation theory shows the vertex must also contain the $\tau_4$-term because only then can the full vertex \emph{Ansatz} reproduce the one-loop result \cite{Chang:2010hb,Davydychev:2000rt}.  Moreover, Eq.\,\eqref{tau4} is precisely the result one obtains on-shell in one-loop Landau-gauge perturbation theory.
%Where a comparison of terms is possible, it is semi-quantitatively in agreement with Refs.\,\cite{Skullerud:2003quBhagwat:2004kj}.

%In the axial-vector and pseudoscalar channels the Ward-Takahashi identity for the Bethe-Salpeter kernel is solved by
In the $0^-$ and $1^+$ channels the Ward-Takahashi identity for the Bethe-Salpeter kernel is solved by
\begin{eqnarray}
\nonumber
2 \Lambda_{5\beta(\mu)} &= & [\tilde \Gamma_{\beta}(q_{+},k_{+})+\gamma_{5}\tilde \Gamma_{\beta}(q_{-},k_{-})
\gamma_{5}]\\
&& \times \frac{1}{S^{-1}(k_{+})+S^{-1}(-k_{-})}\Gamma_{5(\mu)}(k;P)\nonumber\\
\nonumber
&+&\Gamma_{5(\mu)}(q;P)\frac{1}{S^{-1}(-q_{+})+S^{-1}(q_{-})}\\
&& \times [\gamma_{5}\tilde\Gamma_{\beta}(q_{+},k_{+})\gamma_{5}
+\tilde\Gamma_{\beta}(q_{-},k_{-})].
\end{eqnarray}
Given the vertex in Eq.\,\eqref{ourvtx}, we have now completely specified equations in the $0^-$ and $1^+$ channels from which one may obtain bound-state masses and amplitudes.
N.B.\ This solution of the Ward-Takahashi identity for the Bethe-Salpeter kernel is far more general than that presented in Ref.\,\cite{Chang:2009zb}.  Kernels of equal simplicity and power for other channels are readily constructed by analogy.

\begin{figure}[t]
\vspace*{-1ex}

\centerline{\includegraphics[clip,width=0.44\textwidth]{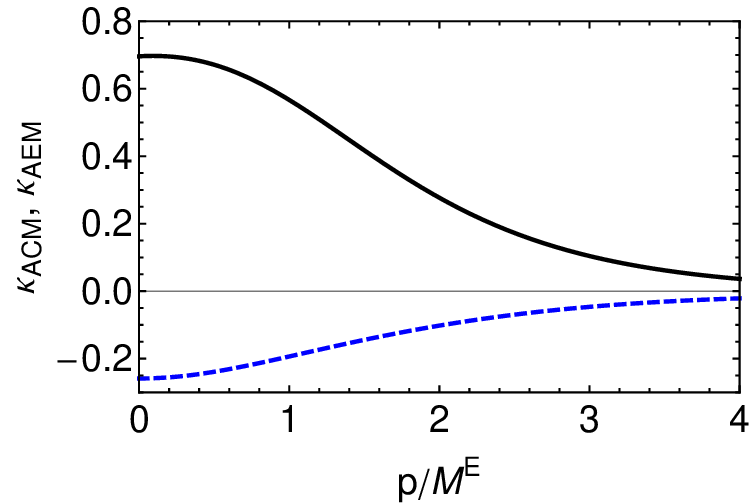}}

\caption{\label{F1}
\emph{Dashed curve} -- Dressed-quark anomalous chromomagnetic moment distribution computed using $\eta=0.65$; and \emph{solid curve} -- Electromagnetic moment distribution computed from the inhomogeneous Bethe-Salpeter equation for the dressed-quark-photon vertex.  In the chiral limit and the absence of DCSB, both curves are zero.  Moreover, absent DCSB-induced terms in the dressed-quark-gluon vertex, Eq.\,\protect\eqref{ourvtx}, both curves would be an order-of-magnitude smaller \protect\cite{Chang:2010hb}.  (The Euclidean constituent-quark mass $M^E := \{p\,|\,p>0,p^2=M^2(p^2)\}=0.35\,$GeV.) }
\end{figure}

\section{Chromo- and Electro-magnetic Moments}
%2 M (dB (-1 + 3 eta) + dA M)
%2 M (0.95 dB + dA M), eta=0.65
Before reporting results for the spectrum it is valuable to illustrate the novel DCSB content of Eq.\,\eqref{ourvtx}.  This may readily be accomplished via a single curve that depicts the dressed-quark anomalous magnetic moment distribution, which is identically zero in the chiral limit \cite{Chang:2010hb}.  One characterizes dressed-quarks through a magnetic moment distribution because a confined quark does not possess a mass-shell \cite{Roberts:1994dr,Roberts:2007ji} and hence one cannot unambiguously assign a single value to its anomalous magnetic moment.

The magnetic moment distribution is computed as follows.  At each value of $p^2$, we define spinors to satisfy a Dirac equation with the constant mass replaced by the running mass: $m\to M(p^2)=:\varsigma$, and use the associated Gordon identity to write
\begin{eqnarray}
\nonumber
\lefteqn{\bar u(p_f;\varsigma) \, \Gamma_\mu( p_f,p_i;k)\,  u(p_i;\varsigma)}\\
& = &
\bar u(p_f) [ F_1(k^2) \gamma_\mu + \frac{1}{2 \varsigma} \,\sigma_{\mu \nu} k_\nu F_2(k^2)] u(p_i).
\label{GenSpinors}
\end{eqnarray}
Now, from Eqs.\,(\ref{ourvtx}) -- (\ref{tau4}), one finds an anomalous chromomagnetic moment distribution
\begin{equation}
\label{kappaacm}
\kappa^{\rm acm}(\varsigma) = \frac{
2 \varsigma (\delta_B^\varsigma (3 \eta - 1) + \varsigma \delta_A^\varsigma)
}
    {\sigma_A^{\varsigma} - 2 \varsigma^2 \delta_A^{\varsigma}+ 2 \varsigma \delta_B^{\varsigma} }\,,
\end{equation}
where $\sigma_A^{\varsigma} = \Sigma_A(\varsigma,\varsigma)$, $\delta_A^{\varsigma} = \Delta_A(\varsigma,\varsigma)$, etc.  In Fig.\,\ref{F1} we depict the result obtain using the interaction in Eq.\,\eqref{CalGQC}.

One computes the dressed-quark anomalous electromagnetic moment distribution from the solution of the inhomogeneous Bethe-Salpeter equation in the color-singlet vector vertex.  Details may be found in Ref.\,\cite{Chang:2010hb}, which also explains the impact of DCSB in the dressed-quark-gluon vertex.  The result is depicted in Fig.\,\ref{F1}.  The marked similarity between this computed result and that produced by the \emph{Ansatz} in Ref.\,\cite{Bashir:2011dp} is noteworthy.

\begin{figure}[t]
\vspace*{-1ex}

\centerline{\includegraphics[clip,width=0.44\textwidth]{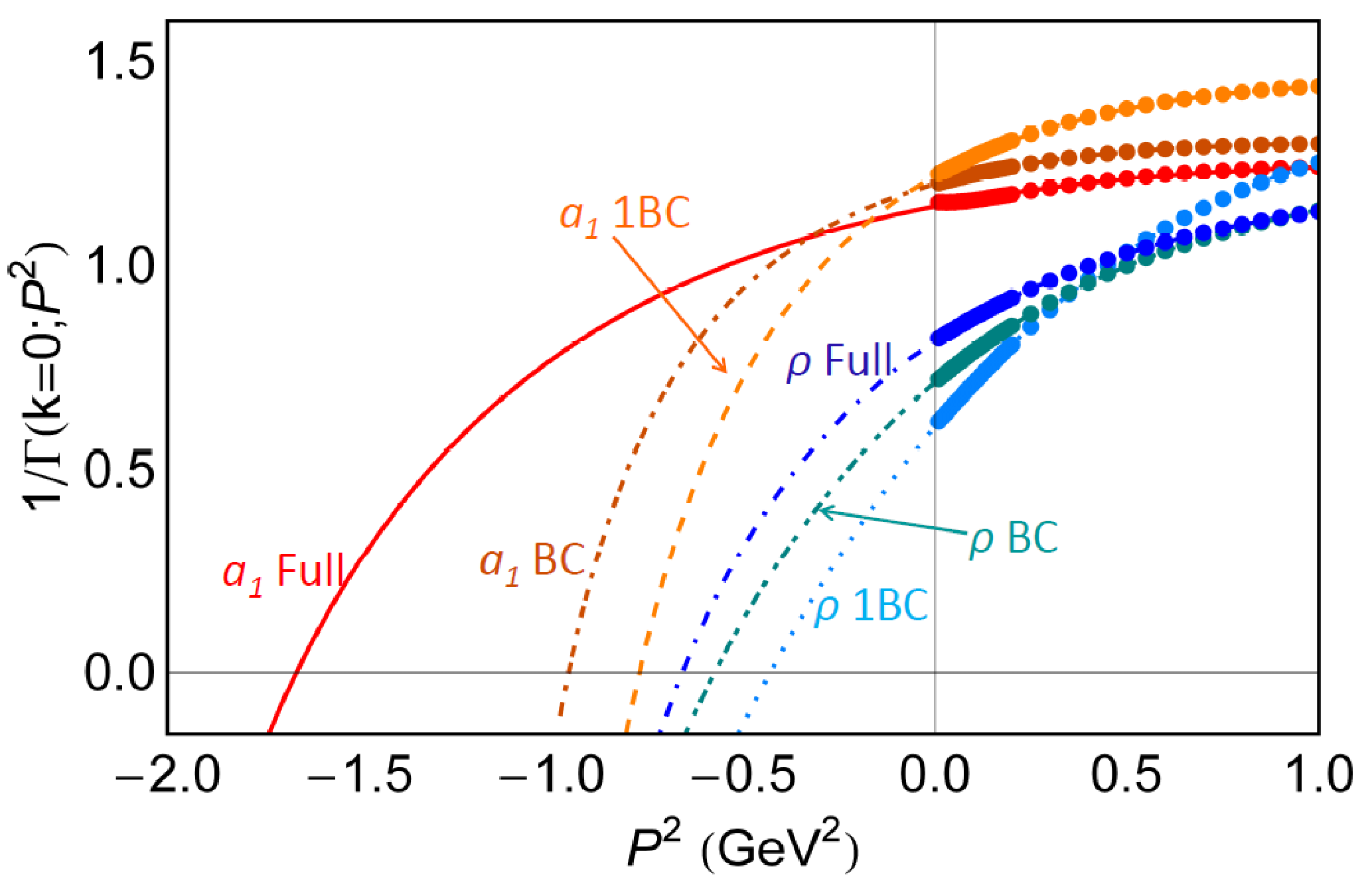}}

\caption{\label{F2} Illustration of the procedure used to determine meson masses.
%, which is analogous to that used in lattice-QCD.
%, which is fully described in Ref.\,\cite{Bhagwat:2007rj} and analogous to the method used in lattice-QCD.
\emph{Solid curve} -- $a_1$-meson, nonperturbative kernel; \emph{dot-dash-dash} -- $a_1$, kernel derived from Eq.\,(\protect\ref{bcvtx}) only (Ball-Chiu, BC); and \emph{dash} -- $a_1$, kernel derived from just the first term in Eq.\,(\ref{bcvtx}) (1BC, a minimal renormalization improvement \protect\cite{Bloch:2002eq} of the leading-order -- RL, rainbow-ladder -- kernel \protect\cite{Bender:1996bb}).
\emph{Dot-dash curve} -- $\rho$-meson, nonperturbative kernel; \emph{Dot-dash-dot} -- $\rho$, BC-kernel; and \emph{dotted} -- $\rho$, 1BC-kernel.
\emph{Points} -- values of $1/\Gamma(k=0;P^2)$ in the given channel computed with the kernel described.  Pad\'e approximants are constructed in each case; and the location of the zero is identified with $(-m_{\rm meson}^2)$.}
\end{figure}

\section{Meson Spectrum}
We compute masses using the method detailed in Ref.\,\cite{Bhagwat:2007rj}, which ensures one need only solve the gap and Bethe-Salpeter equations at spacelike momenta: a significant numerical simplification.  To explain, the inhomogeneous BSE is solved for the complete amplitude in a particular channel on a domain of spacelike total-momenta, $P^2>0$.  Any bound-state in that channel appears as a pole in the solution at $P^2=-m_{\rm meson}^2$.  Denoting the leading Chebyshev moment of the amplitude's dominant Dirac structure by $\Gamma(k;P)$, then $1/\Gamma(k=0;P^2)$ exhibits a zero at $(-m_{\rm meson}^2)$.  The location of that zero is determined via extrapolation of a Pad\'e approximant to the spacelike-behavior of $1/\Gamma(k=0;P^2)$.  This is illustrated for the $\rho$- and $a_1$-channels in Fig.\,\ref{F2}.

Our results are listed in Table~\ref{tablemasses}, wherein
the level of agreement between Cols.~3 and 4 illustrates the efficacy of the method we're using to compute masses: no difference is greater than 1\%.
Next consider $m_\sigma$ and compare Cols.~1--3.  It is an algebraic result that in the RL-truncation of QCD's Dyson-Schwinger equations (DSEs), $m_\sigma \approx 2 M$, where $M$ is a constituent-like quark mass \cite{Roberts:2011cf}.  On the other hand, incorporating the quark mass function into the Bethe-Salpeter kernel via $\Gamma_\mu^{\rm BC}$ generates a strong spin-orbit interaction, which significantly boosts $m_\sigma$ \cite{Chang:2009zb}.  This feature is evidently unaffected by the inclusion of $\Gamma_\mu^{\rm acm}$; i.e., those terms associated with a dressed-quark anomalous chromomagnetic moment.
Since we deliberately omit terms associated with pion final-state interactions in our nonperturbative kernel, it is noteworthy that $m_\sigma$ in Col.~1 matches estimates for the mass of the dressed-quark-core component of the $\sigma$-meson obtained using unitarized chiral perturbation theory \cite{Pelaez:2006nj,RuizdeElvira:2010cs}.

\begin{table}[t]
\caption{\label{tablemasses}
%Computed masses.
%
Col.~1: Spectrum obtained with the full nonperturbative Bethe-Salpeter kernels described herein, which express effects of DCSB: $\omega=0.5\,$GeV; $D\omega=(0.52\,{\rm GeV})^3$; and we assume isospin symmetry, with $m_u^\zeta=m_d^\zeta=m=3.7\,$MeV, $\zeta=19\,$GeV.  The method of Ref.\,\protect\cite{Bhagwat:2007rj} was used: the error reveals the sensitivity to varying the order of Pad\'e approximant.  We compute $f_\pi= 0.091\,$GeV using Eq.\,(15) in Ref.\,\cite{Chang:2009zb}.
Col.~2 -- %(Expt.),
Experimental values; computed, except $m_\sigma$, from isospin mass-squared averages \protect\cite{Nakamura:2010zzi}.
%, which reports the quark-core mass estimated in Refs.\,\ref{Pelaez:2006nj}.
%
Col.~3 -- Masses determined from the inhomogeneous BSE at leading-order in the DSE truncation scheme of Ref.\,\protect\cite{Bender:1996bb} using the interaction in Ref.\,\cite{Chang:2009zb} (with this simple kernel, the Pad\'e error is negligible);
and Col.~4 -- results in Ref.\,\protect\cite{Alkofer:2002bp}, obtained directly from the homogeneous BSE at the same order of truncation.
In the last two cases, $m=5\,$MeV, $\omega=0.5\,$GeV, $D\omega=(0.794\,{\rm GeV})^3$.
}
\begin{center}
% rl tandy & pi & rho & sigma & a1 & b1
% rl Pade
% full
%\begin{tabular}{cccccc}
\begin{tabular*}
{\hsize}
{|l@{\extracolsep{0ptplus1fil}}
|l@{\extracolsep{0ptplus1fil}}
|l@{\extracolsep{0ptplus1fil}}
|l@{\extracolsep{0ptplus1fil}}
|l@{\extracolsep{0ptplus1fil}}|}\hline
\rule{0em}{3ex}
    & This work & Expt.~ & \emph{RL-Pad\'e}~ & \emph{RL-direct}~ \\\hline
$m_\pi$   & $0.138 $~ & 0.138 & 0.138~ & 0.137~ \\
$m_\rho$  & $0.84 \pm 0.03$~ & 0.777 & 0.754~ & 0.758~ \\
$m_\sigma$& $1.13 \pm 0.01$  & $0.4$ -- $1.2$~ & 0.645~ & 0.645~ \\
$m_{a_1}$ & $1.28 \pm 0.01$~ & $1.24 \pm 0.04$~ & 0.938~ & 0.927~  \\
$m_{b_1}$ & $1.24 \pm 0.10$~ & $1.21 \pm 0.02$~ & $0.904 $~ & 0.912~ \\
$m_{a_1}-m_\rho$~ & $0.44 \pm 0.04$ & $0.46 \pm 0.04$ & 0.18 & 0.17 \\
$m_{b_1}-m_\rho$~ & $0.40 \pm 0.14$ & $0.43 \pm 0.02$ & 0.15 & 0.15 \\\hline
\end{tabular*}
%\end{tabular}
\end{center}
\end{table}

Now compare the entries in Rows~2, 4--6.  The $\rho$- and $a_1$-mesons have been known for more than thirty years and are typically judged to be parity-partners; i.e., they would be degenerate if chiral symmetry were manifest in QCD.  Plainly, they are not, being split by roughly $450\,$MeV (i.e., $> m_\rho/2$).  It is suspected that this large splitting owes to DCSB: hitherto, however, no symmetry-preserving bound-state treatment could explain it.  This is illustrated by Cols.~3, 4, which show that whilst a good estimate of $m_\rho$ is readily obtained at leading-order in the systematic DSE truncation scheme of Ref.\,\cite{Bender:1996bb}, the axial-vector masses are much underestimated.  The flaw persists at next-to-leading-order \cite{Watson:2004kd,Fischer:2009jm}.

Our analysis points to a remedy for this longstanding failure.  Using the Poincar\'e-covariant, symmetry preserving formulation of the meson bound-state problem enabled by Ref.\,\cite{Chang:2009zb}, with nonperturbative kernels for the gap and Bethe-Salpeter equations, which incorporate and express effects of DCSB that are impossible to capture in any step-by-step procedure for improving upon the rainbow-ladder truncation, we provide realistic estimates of axial-vector meson masses.
In obtaining these results we found that the vertex \emph{Ansatz} used most widely in studies of DCSB, $\Gamma_\mu^{BC}$, is inadequate as a tool in hadron physics.  Used alone, it increases both $m_\rho$ and $m_{a_1}$ but yields $m_{a_1}-m_\rho=0.22\,$GeV, qualitatively unchanged from the rainbow-ladder-like result (see Fig.\,\ref{F1}).
A good description of axial-vector mesons is achieved by including interactions derived from $\Gamma_\mu^{\rm acm}$; i.e., connected with the dressed-quark anomalous chromomagnetic moment \cite{Chang:2010hb}.  Moreover, used alone, neither term in $\Gamma_\mu^{\rm acm}$ can produce a satisfactory result.  The full vertex \emph{Ansatz} and the associated gap and Bethe-Salpeter kernels described herein are the minimum required.

In preparing Row~5 we obtained additional information.  The leading-covariant in the $b_1$-meson channel is $\gamma_5 k_\mu$.  The appearance of $k_\mu$ suggests that dressed-quark orbital angular momentum will play a significant role in this meson's structure, even more so than in the $a_1$-channel for which the dominant covariant is $\gamma_5\gamma_\mu$.
(NB. In a simple quark-model, constituent spins are parallel within the $a_1$ but antiparallel within the $b_1$.  Constituents of the $b_1$ can therefore become closer, so that spin-orbit repulsion can exert a greater influence.)
This expectation is realized in the result that $m_{b_1}$ is far more sensitive to the interaction's range parameter, $\omega$, than any other state, increasing rapidly with decreasing $\omega$.
Such behavior is readily understood.  The kernel's nonperturbatively-induced spin-orbit interaction acts over a length-scale characterized by $r_c:=1/\omega$.  As $r_c$ is reduced with $D$ fixed, the effective range of spin-orbit repulsion is also reduced, and the mass therefore drops.
%
%omega gets bigger ... broader in momentum space ... narrower in config space ... range reduced over which spin-orbit repulsion can act ... mass gets smaller.
% rho <-> a1: both up-up but L only in a_1.  Mass is SO-repulsion
% b1 <-> a1: b1 up-down, quarks can be closer together cf. a1, so SO-repulsion is enhanced.
%This computation was enabled by validation of the Pad\'e-method for meson mass computation in Ref.\,\cite{Bhagwat:2007rj}, and ...
% Watson, Cassing, Tandy ... other things being equal, saw b1 much more sensitive than a1 to vertex corrections.

\section{Epilogue}
Our results rest on an \emph{Ansatz} for the quark-gluon vertex.  The best available information was used in its construction.  Improvement is nonetheless possible, which will involve elucidating the role of Dirac covariants not yet considered and of resonant contributions; viz., meson loop effects that give widths to some of the states we've considered.  In cases for which empirical width-to-mass ratios are $\lesssim 25$\%, we judge that such contributions can reliably be obtained via bound-state perturbation theory \cite{Pichowsky:1999mu}.  Contemporary studies indicate that these effects reduce bound-state masses but the reduction can uniformly be compensated by a modest inflation of the interaction's mass-scale \cite{Roberts:2011cf,Eichmann:2008ae}, so that the masses in Table~\ref{tablemasses} are semiquantitatively unchanged.  The case of the $\sigma$-meson is more complicated.  However, we predict a large mass for this meson's dressed-quark core, which leaves sufficient room for a strong reduction by resonant contributions \cite{Pelaez:2006nj,RuizdeElvira:2010cs}.
Balance requires we note that one cannot yet completely exclude the possibility of strong couplings between other channels, consequences of which for understanding the spectrum are illustrated in Ref.\,\cite{Nagahiro:2011jn}.

We furnished a continuum framework for computing and explaining the meson spectrum.  It combines a veracious description of pion properties with estimates for masses of light-quark mesons heavier than $m_\rho$.  Our method therefore offers the promise of a first reliable Poincar\'e-invariant, symmetry-preserving computation of the spectrum of light-quark hybrids and exotics; i.e., those putative states which are impossible to construct in a quantum mechanics based upon constituent-quark degrees-of-freedom.  So long as the promise is promptly fulfilled, the approach will provide predictions to guide the forthcoming generation of facilities.

\bigskip

%\begin{acknowledgments}
%
\noindent\textbf{Acknowledgments}. We acknowledge valuable communications with Y.-X.~Liu.
Work supported by
U.\,S.\ Department of Energy, Office of Nuclear Physics, contract no.~DE-AC02-06CH11357.
%\end{acknowledgments}

%\bibliographystyle{../../../../../zProc/z10KITPC/h-physrev4}
%\bibliography{../../../../../CollectedBiB}

\end{document}